\documentclass[%
 reprint,
superscriptaddress,
 nofootinbib,
 amsmath,amssymb,
 aps,
 prl,
twocolumn,
balancelastpage,
]{revtex4-1}
\renewcommand{\vec}[1]{\pmb{#1}}

\usepackage{graphicx}
\usepackage{dcolumn}
\usepackage{bm}
\usepackage{xcolor}
\usepackage{hyperref}
\newcommand{\beq}{\begin{equation}}
\newcommand{\eeq}{\end{equation}}

\begin{document}

\preprint{APS/123-QED}

\title{Magnetic Hair and Reconnection in Black Hole Magnetospheres}

             
\author{Ashley Bransgrove}
\altaffiliation[ashley.bransgrove@columbia.edu]{}
\affiliation{Physics Department and Columbia Astrophysics Laboratory, Columbia University, 538 West 120th Street, New York, NY 10027}

\author{Bart Ripperda}
\affiliation{Center for Computational Astrophysics, Flatiron Institute, 162 Fifth Avenue, New York, NY 10010, USA}\affiliation{Department of Astrophysical Sciences, Peyton Hall, Princeton University, Princeton, NJ 08544, USA}

\author{Alexander Philippov}
\affiliation{Center for Computational Astrophysics, Flatiron Institute, 162 Fifth Avenue, New York, NY 10010, USA}

\begin{abstract}
The no-hair theorem of general relativity states that isolated black holes are characterized by three parameters: mass, spin, and charge. In this Letter we consider Kerr black holes endowed with highly magnetized plasma-filled magnetospheres. Using general relativistic kinetic plasma and resistive magnetohydrodynamics simulations, we show that a dipole magnetic field on the event horizon opens into a split-monopole and reconnects in a plasmoid-unstable current-sheet. The no-hair theorem is satisfied, in the sense that all components of the stress-energy tensor decay exponentially in time. We measure the decay time of magnetic flux on the event horizon for plasmoid-dominated reconnection in collisionless and collisional plasma. The reconnecting magnetosphere should be a powerful source of hard X-ray emission when the magnetic field is strong.  
\end{abstract}

\maketitle
Black holes (BHs) formed by the collapse of a magnetized progenitor are born with magnetic fields penetrating the event horizon. There are several possible scenarios, such as the spin-down of a rotationally supported hyper-massive neutron star (NS) \citep{Falcke_2014}, or gravitational collapse induced by the accretion of dark matter onto the NS core \citep{Goldman_1989}. BHs can also acquire magnetic flux later in life by merging with a magnetized NS \cite{east2021multimessenger}, or in accretion flows. The fate of the magnetic flux (hair) on the event horizon should be in accordance with the no-hair theorem of general relativity. 

The original no-hair conjecture \citep{MTW} states that all stationary, asymptotically flat BH spacetimes should be completely described by the mass, angular momentum, and electric charge. It was later proved formally that any field with zero rest mass and arbitrary integer spin is radiated away on a light crossing timescale \citep{Price_1972b}. In particular, the multipole component $l$ of a magnetic field in vacuum decays as $t^{-(2l+2)}$. However, magnetized BHs are unlikely to exist in vacuum. If a BH is formed by the collapse of a magnetized star, plasma will inevitably exist in the magnetosphere around the newly formed event horizon. Furthermore, BHs can generate a self-regulated plasma supply through electron-positron discharges near the event horizon \citep{BZ_1977,Parfrey_2019,Crinquand_2020}. The discharges can fill the magnetosphere with plasma in a light crossing time.

The presence of highly conducting plasma, and thus non-zero stress-energy tensor of matter, dramatically changes the vacuum dynamics assumed in the classical no-hair theorem. Essentially, in the limit of vanishing resistivity a topological constraint is imposed which prevents the magnetic field from sliding off the event horizon \cite{lyutikov_slowly_2011}. The only way for the BH to lose its magnetic field is for the field to change its topology (reconnect). Fast magnetic reconnection occurs through the tearing instability \cite{Bhattacharjee_2009}. A chain of plasmoids (magnetic loops containing plasma) forms along the reconnection layer which are ejected at relativistic velocities. For highly magnetized collisionless plasma (as expected in a BH magnetosphere), the reconnection rate $v_\text{rec}\sim 0.1 c$ is independent of the magnetization \cite{Sironi_2014,Guo_2014,Werner_2015}. The lifetime of the magnetic flux on the event horizon should be determined in part by this universal reconnection rate. 

Previous work in an ideal fluid approximation correctly established the qualitative evolution of a dipole magnetic field on the event horizon opening into a split-monopole \cite{lyutikov_slowly_2011}. However, it neglected collisionless physics, and  was performed at low numerical resolution such that the reconnection was not in the high Lundquist number regime \cite{lyutikov_slowly_2011,Lehner_2012}. This lead to the conclusion of an extremely long lifetime of the magnetic flux on the event horizon, dictated by the resistive timescale of the plasma \cite{lyutikov_slowly_2011}. In this Letter we describe for the first time GRPIC (general-relativistic particle-in-cell) and GRRMHD (general-relativistic resistive magnetohydrodynamics) simulations which are converged and produce the correct reconnection physics.     

The system is solved numerically in Kerr spacetime. Kerr-schild coordinates ($t,r,\theta,\phi$) are used so that all quantities are regular at the event horizon. The dimensionless BH spin is set to $a=0.99$ to maximize the ergosphere volume. We define ``fiducial observers" (FIDOs), whose worldlines are normal to spatial hypersurfaces. 
We assume that the NS was already surrounded by plasma, and that it collapsed into a BH before the simulation begins. This setup is sufficient to test the no-hair theorem because when plasma is present, the magnetic field cannot escape before the event horizon has formed \cite{lyutikov_slowly_2011}. The initial condition for all simulations is a magnetic dipole described by the vector potential $ A_{\phi} = B_0 \sin^2\theta / r$, where $B_0$ is the dimensionless magnetic field strength at the horizon as measured by the FIDO. The magnetic field components are obtained from $B^i = \epsilon^{ijk}\partial_j A_k / \sqrt{\gamma}$, where $\sqrt{\gamma}$ is the spatial metric determinant. In vacuum non-zero $\nabla \times (\alpha \vec{B})$ is quickly radiated away or swallowed by the BH ($\alpha$ is the lapse). However, when plasma is present non-zero $\nabla \times (\alpha \vec{B})$ drives currents which slow down the balding process.

The kinetic plasma simulations are performed using the general-relativistic particle-in-cell (PIC) code {\tt{Zeltron}} \cite{Parfrey_2019}. We solve the equations of motion for pair plasma particles, together with Maxwell’s equations for electromagnetic fields. All lengths are given in units of $r_g=GM/c^2$ with $M$ the BH mass, and times in units of $r_g/c$. The particles have mass $m$, and charge $\pm e$. The GRPIC simulations begin with vacuum, and plasma particles are injected with density proportional to the local parallel electric field as a proxy for the electron-positron discharge (see \cite{Parfrey_2019} for details of the injection scheme).  

We set the dimensionless magnetic field strength at the event horizon $B_0=r_g/r_L$, with $r_L$ the Larmor radius. For the gravitational collapse of a NS it implies $B_0 \sim  10^{14} (M/M_\odot) (B/10^{12}~G)$. In this work we scale it down, and consider $B_0 \sim10^{4}, 3\times10^{4}, 10^{5}$ (Table~\ref{table}). We show that our results are independent of $B_0$, as long as the plasma is highly magnetized. The characteristic minimum plasma density required to support the rotating magnetosphere is the Goldreich-Julian number density \cite{goldreich_pulsar_1969}, $n_\text{0}=\Omega_H B_0/(2\pi ce)$, where $\Omega_H=a c r_g/[r_H^2 + (r_g a)^2]$ is the angular velocity of the event horizon radius $r_H = r_g(1+\sqrt{1-a^2})$. It implies the characteristic magnetization $\sigma_0 = B_0^2 / (4\pi n_{0} m c^2) = (1/2)(\omega_{B} / \Omega_H) = (1/4)(\omega_p / \Omega_H)^2\gg 1$, where $\omega_{p} = (4\pi n_\text{0}e^2 / m)^{1/2}$ is the plasma frequency and $\omega_B=c/r_L$ the Larmor frequency. We have preserved the astrophysically relevant hierarchy of scales $r_{L} \ll \lambda_{p} \ll r_g$, and $\Omega_H \ll \omega_p \ll \omega_{B}$, where $\lambda_p = c / \omega_{p}$ is the plasma skin depth.

\begin{table}[h]
\centering
\caption{Summary of the simulation parameters. For all GRRMHD runs the diffusivity is $\eta=10^{-5}$. For MHD runs $N_r\times N_\theta \times N_{\phi}$ refers to the effective resolution. Runs with $N_\phi=1$ are axisymmetric, while those with $N_\phi>1$ refer to 3D simulations. All models have spin $a=0.99$ except VAC0, which has $a=0$. }
\begin{tabular}{c c c c}
\hline
Model  & $r_L$         & $\lambda_p$ & $N_r\times N_\theta \times N_\phi$  \\
\hline
VAC0 & ---  & --- & $9600 \times 8016 \times 1$  \\
VAC1 & ---  & --- & $9600 \times 8016 \times 1$  \\
GRPIC1 & $1\times10^{-5}$  & $3\times10^{-3}$ & $2880 \times 2160 \times 1$  \\
GRPIC2 & $3\times 10^{-5}$ & $6\times10^{-3}$ & $2880 \times 2160 \times 1$  \\
GRPIC3 & $1\times10^{-4}$  & $1\times10^{-2}$ & $2880 \times 2160 \times 1$  \\
GRRMHD1 & ---  & --- & $6144 \times 3072 \times 1$  \\
GRRMHD2 & ---  & --- &   $ 3072 \times 1536 \times 1536 $ \\
\hline
\end{tabular}
\label{table}
\end{table}

The computational domain of the axisymmetric GRPIC simulations covers $0.99 \leq r \leq 75$, and $0\leq \theta \leq \pi$. Simulations for each of the (3) magnetic field strengths were performed at two resolutions to check for numerical convergence (a total of 6 kinetic plasma simulations): (i) $N_r\times N_\theta = 1440 \times 1080$, and (ii) $N_r\times N_\theta = 2880 \times 2160$. The grid is uniformly spaced in $\log r$ and $\cos\theta$, so that resolution is concentrated near the BH horizon, and the equator. We check that the plasma skin depth is well resolved a posteriori, since the plasma density is determined self-consistently. Electromagnetic fields are damped and particles are absorbed at the outer boundary in order to mimic an outflow boundary condition. For $r \leq r_H$ all characteristics are inward, and causality prevents waves and plasma from escaping. Therefore, the equations are solved without modification at the event horizon, and no boundary condition is imposed there.

The GRRMHD simulations are performed using the Black Hole Accretion Code \cite{BHAC,Olivares2019,ripperda2019b}. A minimum density is set throughout the domain such that the magnetization $\sigma\gg 1$, and the plasma is nearly force-free. We set a constant and uniform diffusivity $\eta=10^{-5}$, so that the Lundquist number $S = v_A L / \eta \approx \eta^{-1} = 10^5$ is above the plasmoid instability limit $S>10^4$ \cite{Bhattacharjee_2009}, where $v_A\approx c$ is the Alfv\'en speed and $L\approx r_e - r_g \approx 1$ is the characteristic length of the current-sheet inside the ergosphere.  

The computational domain of the GRRMHD simulations covers $0.99 \leq r \leq 200$, $0\leq \theta \leq \pi$, and $0\leq \phi \leq 2\pi$. By adding AMR, we increase resolution at the current-sheet to assure convergence. The base grid, and additional AMR blocks are uniformly spaced in $\log r$, and $\phi$, while the $\theta$ grid is concentrated near the equator.

\begin{figure*}
\centering
\includegraphics[width=\textwidth]{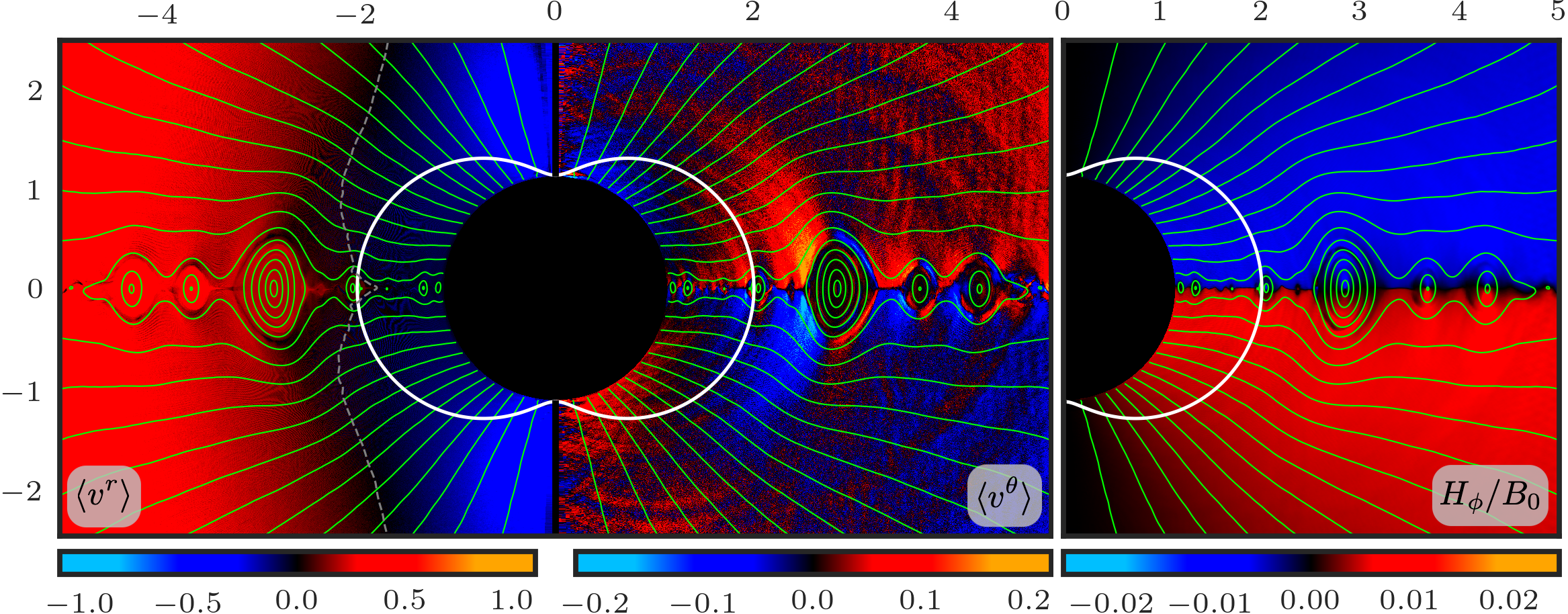} 
\caption{
Reconnecting magnetosphere in the FIDO frame (GRPIC1) at $t=100$~$r_g/c$. Green curves show poloidal magnetic flux surfaces, and white curves show the boundary of the ergosphere. The black circle is the interior of the BH event horizon. Left: Color shows radial and $\theta$ components of the bulk plasma 3-velocity in the orthonormal tetrad basis. The grey dashed curve indicates the stagnation surface defined by $\langle v^r \rangle =0$. Right: Azimuthal component of the auxiliary field $\vec{H}$.}
\label{bulk_v}
\end{figure*}

The evolution of all simulations is qualitatively similar. In GRPIC simulations, strong electric fields induced by spacetime rotation near the event horizon trigger particle injection which rapidly fills the magnetosphere with plasma up to a density $n \sim \mathcal{M} n_{0}$, where $\mathcal{M}\sim$~few is the multiplicity, while GRRMHD simulations begins with a static low-density plasma throughout the domain. In the ergosphere plasma is dragged into co-rotation with the BH, bending field-lines in the $\phi$ direction and inflating the poloidal magnetic field. As field-lines extend in the radial direction, flux on the horizon moves toward the equator, and some loops which close inside the ergosphere are pushed into the BH. After $t \approx 40$ $r_g/c$, the dipole has opened into a split-monopole with $\vec{\Omega}\cdot\vec{B}_p>0$ in both hemispheres, where $\vec{\Omega}$ is the angular velocity vector of the BH and $\vec{B}_p$ is the poloidal magnetic field. The field-lines rotate rigidly with angular velocity $\Omega_F=\Omega_H/2$, in agreement with force-free solutions \cite{BZ_1977}. The toroidal magnetic field $H_\phi$ has opposite sign to $B^r$ in each hemisphere (Fig.~\ref{bulk_v}, right) indicating swept-back field-lines, where $\vec{H}=\alpha\vec{B}-\vec{\beta}\times\vec{D}$, $\vec{\beta}$ is the shift, and $\vec{D}$ the electric field. A well defined MHD stagnation surface is established, separating regions of inflow $\langle v^r\rangle <0$, and outflow $\langle v^r\rangle >0$ (Fig.~\ref{bulk_v}, dashed grey curve). Here $\langle ... \rangle$ indicates averaging over the particles in a single grid-cell.

Magnetic reconnection is first triggered near the stagnation surface in both GRPIC and GRRMHD, and rapidly spreads along the entire current-sheet. The onset of reconnection occurs later in GRRMHD $t\sim 70$ $r_g/c$, compared to GRPIC $t\sim 30$ $r_g/c$. However, once the current-sheet is sufficiently thin the tearing instability develops and a chain of self-similar plasmoids forms. Generally plasmoids born inside the stagnation surface move slowly ($v<0.1c$) toward the event horizon and fall into the BH, while those born outside are ejected from the magnetosphere and accelerate to relativistic velocities ($v\approx c$). Therefore, we identify the stagnation surface at the equator as a main site of field-line ``pinching", and a primary X-point in the global magnetosphere. Occasionally plasmoids born inside the stagnation surface have sufficient kinetic energy to escape. 

We analyzed the reconnection rate for all simulations by measuring the inflow velocity of flux into the current-sheet. The analysis is performed by transforming the electric and magnetic field components into the locally Minkowski reference frame of the FIDO. The inflow velocity is then calculated using the component of $\vec{E}\times\vec{B}$ in the direction perpendicular to the current-sheet, and avoiding plasmoids. We confirm $\sigma\gg 1$ in the upstream plasma, so that $v_A/c = (\sigma / (\sigma+1))^{1/2}\approx 1$, and the reconnection is in the relativistic regime. All components of the magnetic field change sign at the current-sheet, indicating zero guide-field reconnection.

\begin{figure*}
\centering
\includegraphics[width=\textwidth]{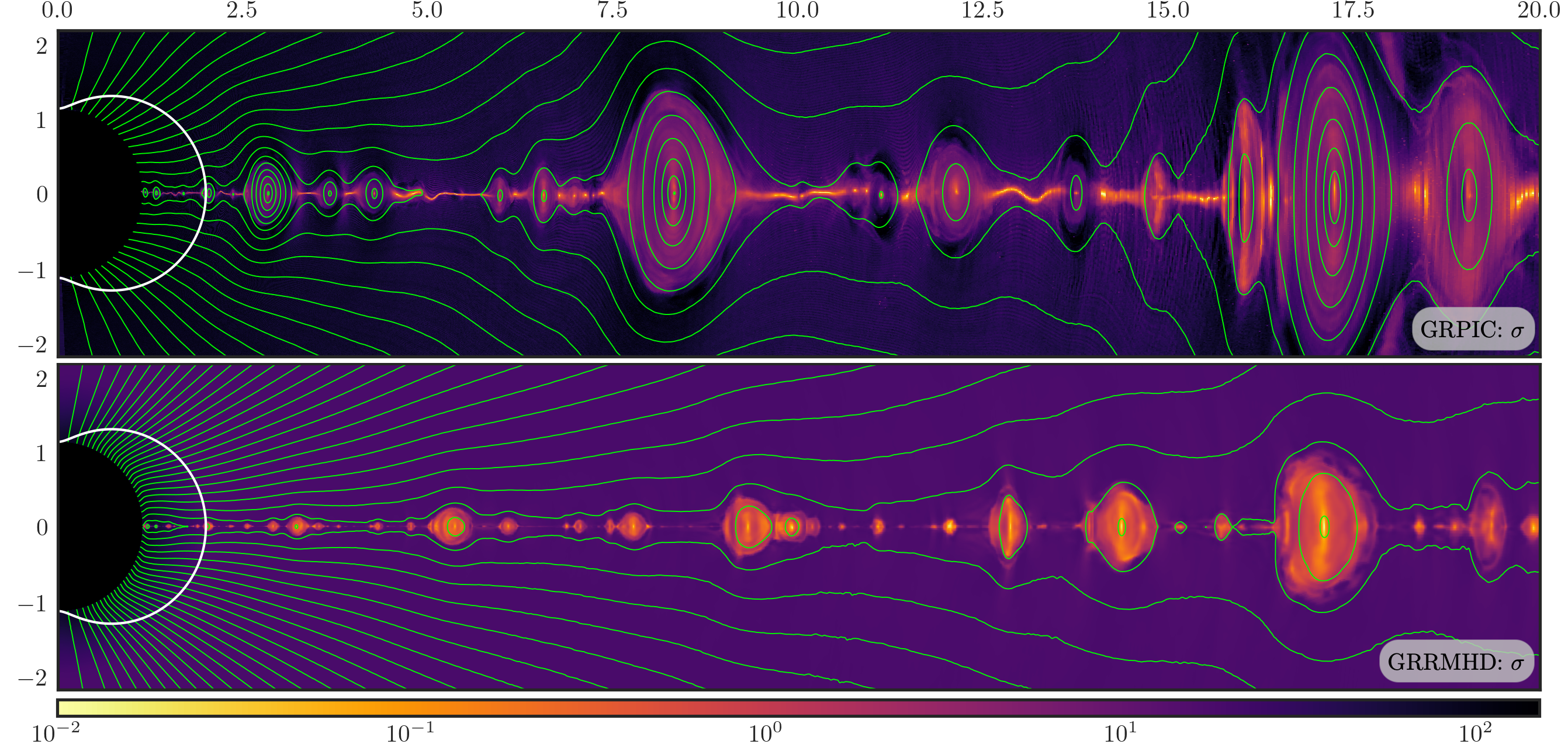} 
    \caption{Different realizations of the reconnecting magnetosphere in the FIDO frame. Color shows the cold plasma magnetization $\sigma$. Top: GRPIC1 at $t=100$~$r_g/c$, $\sigma=B^2/(4\pi m n c^2)$. Bottom: GRRMHD1 at $t=311$~$r_g/c$, $\sigma=B^2/(4\pi \rho c^2)$. The GRPIC simulation (top) displays larger plasmoids than GRRMHD (bottom) due to the faster reconnection rate.}
    \label{scales}
\end{figure*}

The measured reconnection rate in the GRPIC simulations $v_\text{rec}\approx0.1c$ is consistent with studies of magnetic reconnection in relativistic collisionless plasmas \cite{Sironi_2014,Guo_2014,Werner_2015}. For the GRRMHD simulations the high Lundquist number $S\approx 10^5\gg 10^4$ ensures that the reconnection occurs deep in the plasmoid dominated regime \cite{Ripperda_2020}. The reconnection rate in resistive MHD at high Lundquist number is $v_\text{rec}\approx0.01 v_A$ \cite{Bhattacharjee_2009}, which is confirmed by our measured $v_\text{rec}\approx 0.01c-0.02c$ and is consistent with other studies in relativistic MHD \cite{ripperda2019} where the reconnection dynamics is modified by $v_A\rightarrow c$. In GRPIC simulations the plasmoids grow at a rate $\sim 0.1c$, until they are ejected and the growth is suppressed as they reach relativistic velocities. Thus the plasmoids are on average smaller in GRRMHD simulations (Fig.~\ref{scales}), where the growth rate $\sim 0.01c$ is smaller.

Reconnection in collisionless pair plasma occurs due to kinetic effects resulting from the divergence of the anisotropic electron pressure tensor, which plays the role of an effective non-uniform diffusivity \cite{Bessho2005}. Therefore, the difference in reconnnection rates between the two formalisms can be attributed to the use of a uniform diffusivity in GRRMHD as a proxy for kinetic effects, representing the simplest model of reconnection and plasmoid formation, while in GRPIC the dissipation at the current-sheet is determined from first principles. 

The reconnection is collisionless when the plasma skin depth $\lambda_{p}$ is larger than the elementary current-sheet width in the resistive-MHD chain $w\sim 100 \eta/v_A\sim 100 \eta/c$ \cite{Bhattacharjee_2009,Uzdensky_2010}, where $\eta$ is the diffusivity due to coulomb collisions of pairs. Since our simulations do not include the detailed pair production and collision physics, we estimate analytically when this condition is satisfied (see supplement). The temperature of the reconnection layer is estimated by assuming the combined pressure of radiation and pairs is comparable to $B^2/(8\pi)$. The density of pairs is then given by the annihilation balance. We find that the reconnection is evidently collisionless when $B\ll 10^{12}$~G. However, if the magnetic field is very strong $B\gtrsim 10^{12}$~G, or pair production is very efficient, the separation between the two regimes is less clear, and a self consistent calculation is required to determine the reconnection rate. However, even in this intermediate case, the GRRMHD simulations described in this work with uniform $\eta$ provide a lower limit on the reconnection rate.

\begin{figure}
\centering
\includegraphics[width=0.47\textwidth]{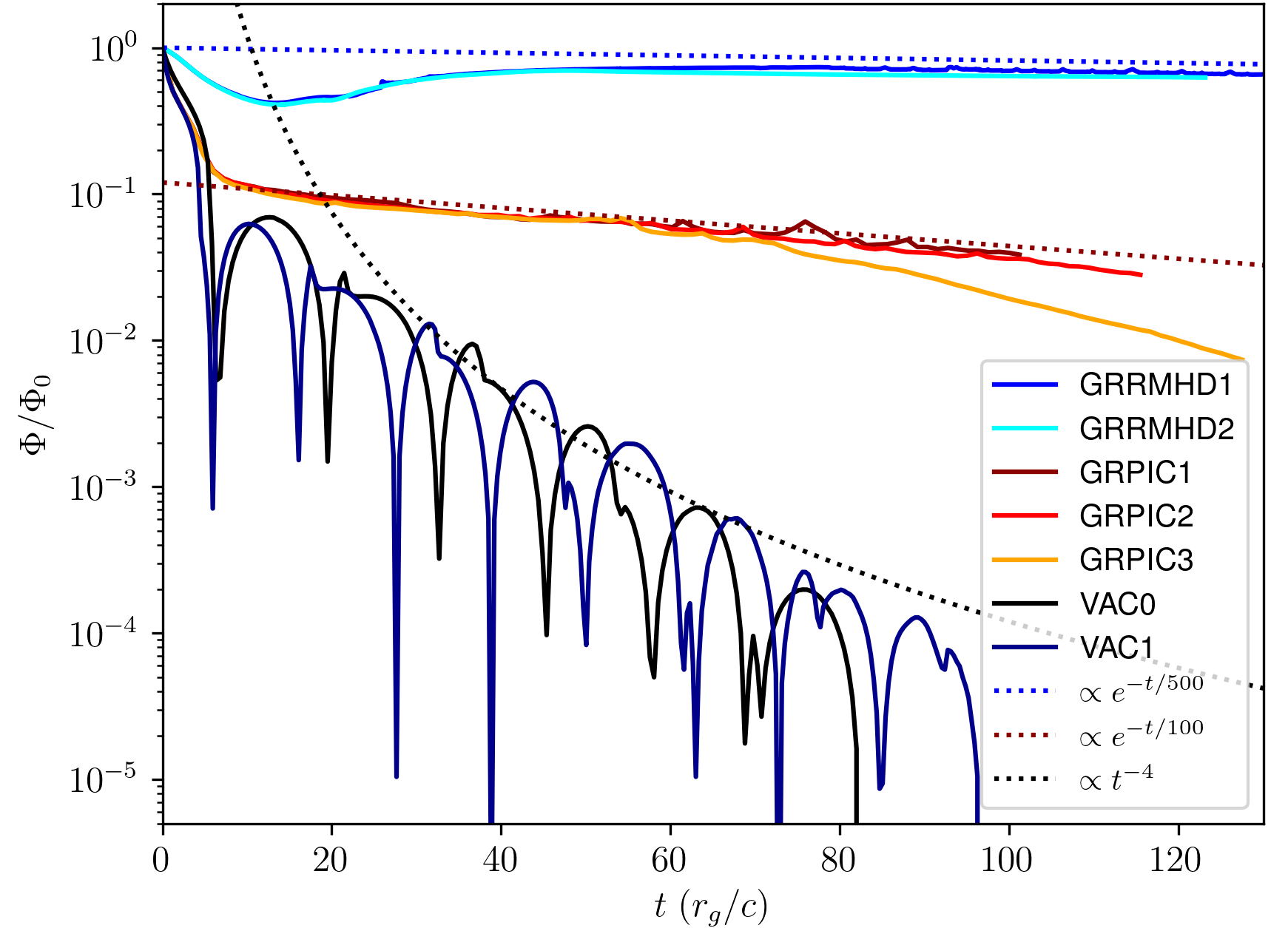} 
\caption{Flux on the event horizon vs time for vacuum (power law decay), collisional MHD plasma (exponential decay), and collisionless plasma (faster exponential decay).}
\label{flux}
\end{figure}

The magnetic flux on the event horizon $\Phi$ decays quasi-exponentially with time (Fig.~\ref{flux}). In GRPIC simulations the flux decays with characteristic timescale $\tau\approx 100$~$r_g/c$, and in GRRMHD simulations $\tau\approx 500$~$r_g/c$ (Fig.~\ref{flux}). The difference in timescales can be attributed to different reconnection rates in these formalisms, which differ by a factor $\sim 5$. 
Since $B$ and $n\propto B$ decay exponentially, all components of the stress-energy tensor become vanishingly small at late times and the no-hair theorem is satisfied. We calculate the charge of the BH at the end of the GRPIC simulation as $Q = (1/4\pi)\int D^r \sqrt{\gamma}\text{d}\theta\text{d}\phi$ at $r=r_H$ and find that $Q=0$, so the final state is a Kerr BH. The decay timescale converges with decreasing $r_L/r_H$ in GRPIC simulation (Fig.~\ref{flux}), indicating the correct asymptotic behaviour with a sufficient separation of scales. Therefore, the measured decay timescale is independent of $B$, as long as the plasma is highly magnetized, $\sigma \gg 1$, and finite Larmor radius corrections are negligible, $r_L/r_g \ll 1$.

The evolution of $\Phi$ is estimated analytically using Faraday's law, and assuming a constant reconnection rate 
on the equator at the stagnation surface (see supplement) \cite{crinquand2020synthetic}. In this toy model $\Phi$ decays exponentially on a timescale $\tau \approx 3 r_g /  \langle v^\theta \rangle$, with $\langle v^\theta \rangle$ the $\theta$ component of the plasma 3-velocity in the orthonormal tetrad basis  (Fig.~\ref{bulk_v}). 
For example, in GRPIC1 $\langle v^\theta \rangle\approx 0.02-0.04c$ at the current-sheet implies $\tau\sim 100$~$r_g / c$, consistent with Fig.~\ref{flux}. The local reconnection rate observed by the FIDO is estimated by taking into account time dilation at the stagnation surface. For GRPIC1 it implies $\langle v^\theta \rangle/(c\alpha) \sim 0.05$, with $\alpha$ evaluated on the equator at the stagnation surface, consistent with the measured values. 

\begin{figure*}
\centering
\includegraphics[width=\textwidth]{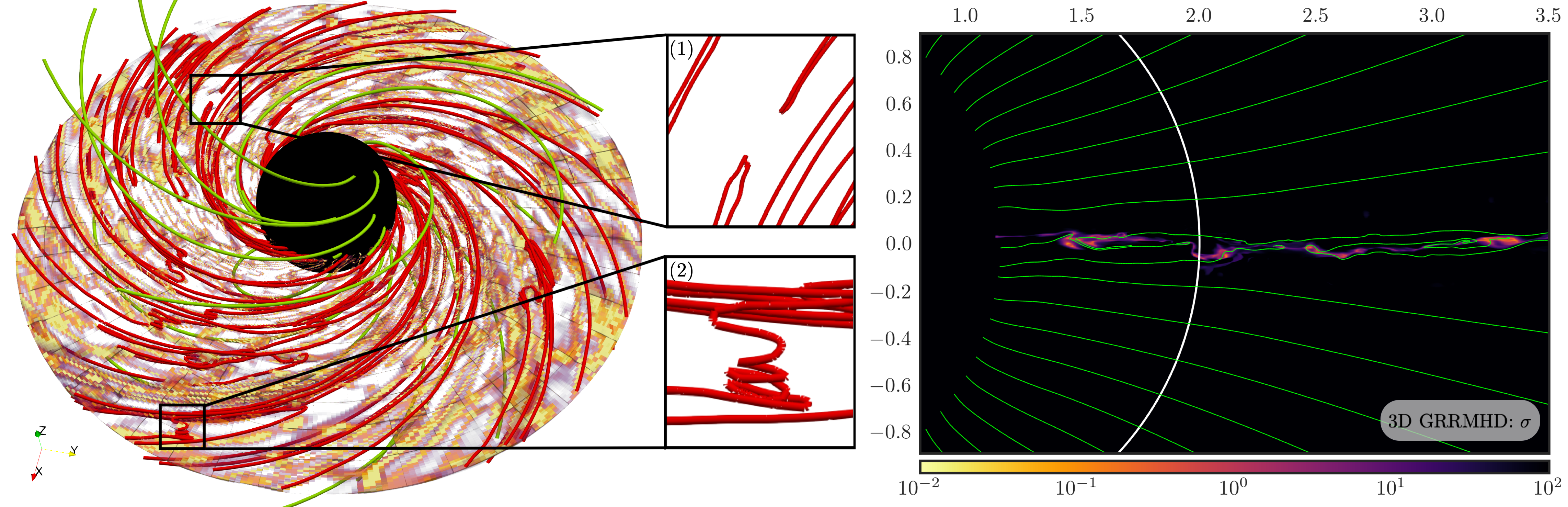} 
    \caption{Reconnecting magnetosphere in 3D (GRRMHD2) at $t=118$~$r_g/c$. Left: Volume rendering shows $\sigma=B^2/(4\pi \rho c^2)$
    , green tubes are magnetic field-lines which penetrate the event horizon, red tubes are magnetic field-lines which are reconnecting in the current-sheet. Right: 2D slice of GRRMHD2 in the $\phi=0$ half-plane. Color shows $\sigma$, green curves are magnetic field-lines in the $\phi=0$ half-plane. The picture highlights the non-axisymmetric nature of reconnection in 3D, yet still displays similar fundamental structures ---  X-points (inset 1), and helical winding of magnetic field-lines in plasmoids (flux ropes) (inset 2).}
    \label{3D}
\end{figure*}

In  3D (GRRMHD2), the balding proceeds similarly to the axisymmetric simulations (Fig.~\ref{flux}, cyan curve), but the plasmoid instability leads to non-axisymmetric (in $\phi$) structures.  Therefore, 3D plasmoids, or flux tubes of tangled field-lines with a finite extent in $\phi$, generally display more complex topologies than those in 2D (Fig.~\ref{3D}).

The flux of conserved energy through spherical shells, as seen by an observer at infinity is comparable in magnitude to $L_\text{BZ}=0.053\Omega_{H}^2\Phi^2/(4\pi c)$ \cite{Tchekhovskoy_2011}, indicating successful activation of the Blandford-Znajek mechanism (see supplement) \cite{BZ_1977}. Large fluctuations up to several $L_\text{BZ}$ are seen at the locations of plasmoids. We observe the emission of fast modes from plasmoid mergers (Fig.~\ref{scales}). In the high$-\sigma$ limit, and where $B\sim 10^6$~G, these fast modes correspond to vacuum electromagnetic waves in the radio band, and could be observed as coherent radio emission \cite{Philippov_2019}. The escaping giant plasmoids (Fig.~\ref{scales}) may shock the upstream wind, resulting in coherent synchrotron maser emission \cite{Lyubarsky_2014,Beloborodov_2017b}. For collisionless plasma, we measure the total dissipative power as seen by an observer at infinity $L_{\text{diss},\infty}\approx 0.4 L_\text{BZ}$. When the magnetic field is strong ($B\gtrsim 10^6$~G) as expected in BH-NS mergers, the reconnection is radiative and most of the dissipated magnetic energy will go into photons. In this regime, $L_{\text{diss},\infty}\approx 0.4 L_\text{BZ} \sim 4\times 10^{45} M_{10\odot}^2 B_{12}^2$~erg~s$^{-1}$ corresponds to emission in the hard X-ray band \cite{Beloborodov_2020}.  We also observe a population of negative energy-at-infinity particles localized in the current sheet inside the ergosphere. They contribute to $\vec{J} = (c/4\pi) \nabla\times \vec{H}$, and some are advected into the BH with plasmoids --- an instance of the Penrose process facilitated by magnetic reconnection \cite{Parfrey_2019,Comisso_2021}.

We considered Kerr BH's endowed with highly magnetized plasma-filled magnetospheres. We find that: (i) The no-hair theorem holds, in the sense that all components of the stress-energy tensor decay exponentially in time, (ii) Reconnection occurs at the universal rate when measured in the locally Minkowski frame of the FIDO, (iii) The lifetime of the magnetic field on the event horizon is controlled by the local reconnection rate measured by the FIDO in concert with other global effects, and (iv) The final state is a Kerr BH with charge $Q=0$. Balding BHs resulting from the merger or collapse of compact objects should appear as a spectacular source of hard X-rays for a short duration, similar to the flares of galactic magnetars. Observation of the X-rays requires a clean environment around the BH. It is possible during the gravitational collapse of a rotationally supported NS, and in BH-NS mergers with a high mass ratio, so that the NS falls through the event horizon without forming a torus or disk. Gamma-ray bursts and other collapsars may be different to the scenario described in this work, depending on how much matter surrounds the newly formed BH. The decay of magnetic flux on the event horizon may also explain powerful X-ray and near-infrared flares and hot spots \cite{Gravity2018} driven by plasmoid-regulated reconnection in magnetically dominated supermassive BH magnetospheres \cite{Ripperda_2020,Comisso_2021}. The faster reconnection rate in collisionless plasma implies that larger plasmoids, powering a flare near the BH, can form in a shorter time and in this way regulate the typical flare duration. 

\begin{acknowledgments}
The authors thank Y. Levin and M. Medvedev for useful discussions. A.B. acknowledges the 2019 Summer School at the Center for Computational Astrophysics, Flatiron Institute, and a 2020 Flatiron Institute Summer Internship where part of this work was completed. B.R. is supported by a Joint Princeton/Flatiron Postdoctoral Fellowship. A.P. acknowledges support by the National Science Foundation under Grants No.  PHY-2010145. Research at the Flatiron Institute is supported by the Simons Foundation. The computational resources used in this work were provided by facilities supported by the Scientific Computing Core at the Flatiron Institute; and by the VSC (Flemish Supercomputer Center), funded by the Research Foundation Flanders (FWO) and the Flemish Government – department EWI. This research is part of the Frontera computing project at the Texas Advanced Computing Center (LRAC-AST21006). Frontera is made possible by National Science Foundation award OAC-1818253.

\end{acknowledgments}

\onecolumngrid

\section{Supplemental Materials}
In this supplemental material we provide additional details about numerical convergence, collisionality of the reconnection layer, the flux decay timescale, and the luminosity.

\section{Numerical Convergence}

\begin{figure}[h!]
\centering
\includegraphics[width=.44\textwidth]{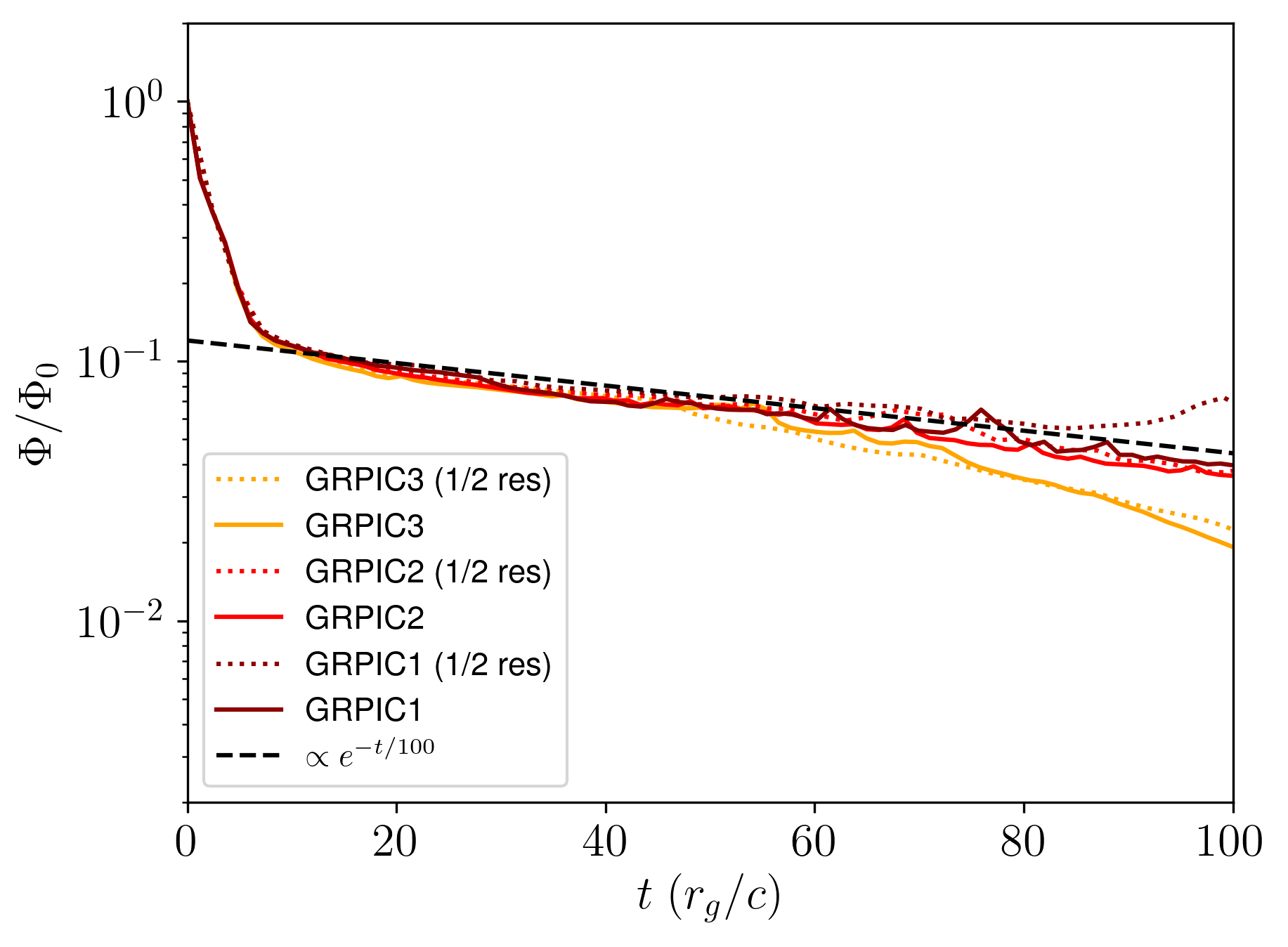} 
\includegraphics[width=.44\textwidth]{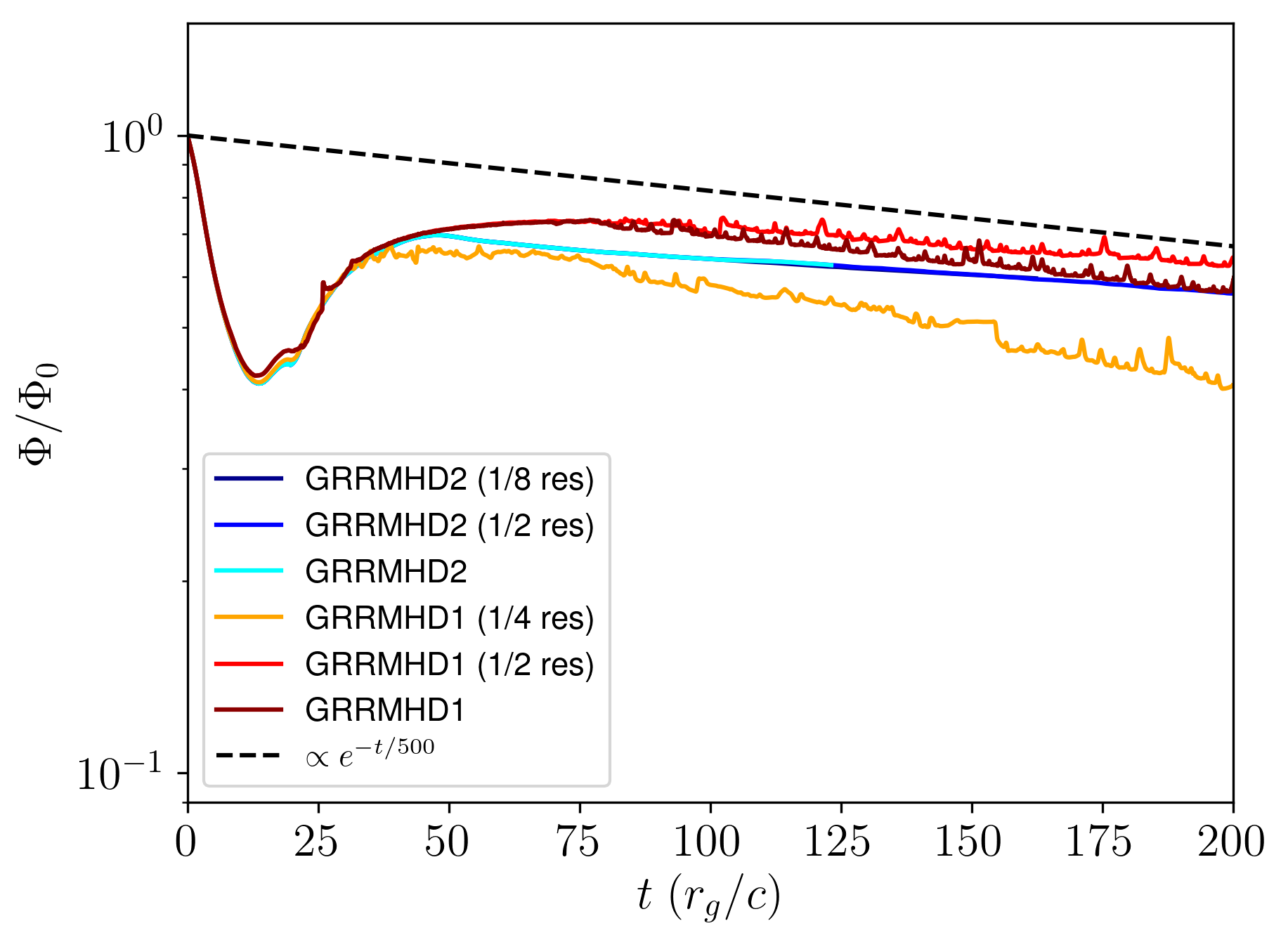} 
\caption{Convergence study for each simulation. Left: Flux on the event horizon vs time at 2 resolutions for each of the simulations GRPIC1, GRPIC2, and GRPIC3. Right: Flux on the event horizon vs time at 3 resolutions for simulation GRRMHD1 and GRRMHD2.}
\label{convergence}
\end{figure}

Figure~\ref{convergence} shows time evolution of the magnetic flux on the event horizon vs time at different grid resolutions. In GRPIC the simulations are well converged at all scale separations for the resolution $2880\times2160$. In GRRMHD the axisymmetric simulation is well converged, resolving the current layer by approximately 10 cells over its width for $\eta=10^{-5}$ at a resolution $6144 \times 3072$.

\begin{figure}[h!]
\centering
\includegraphics[width=0.50\textwidth]{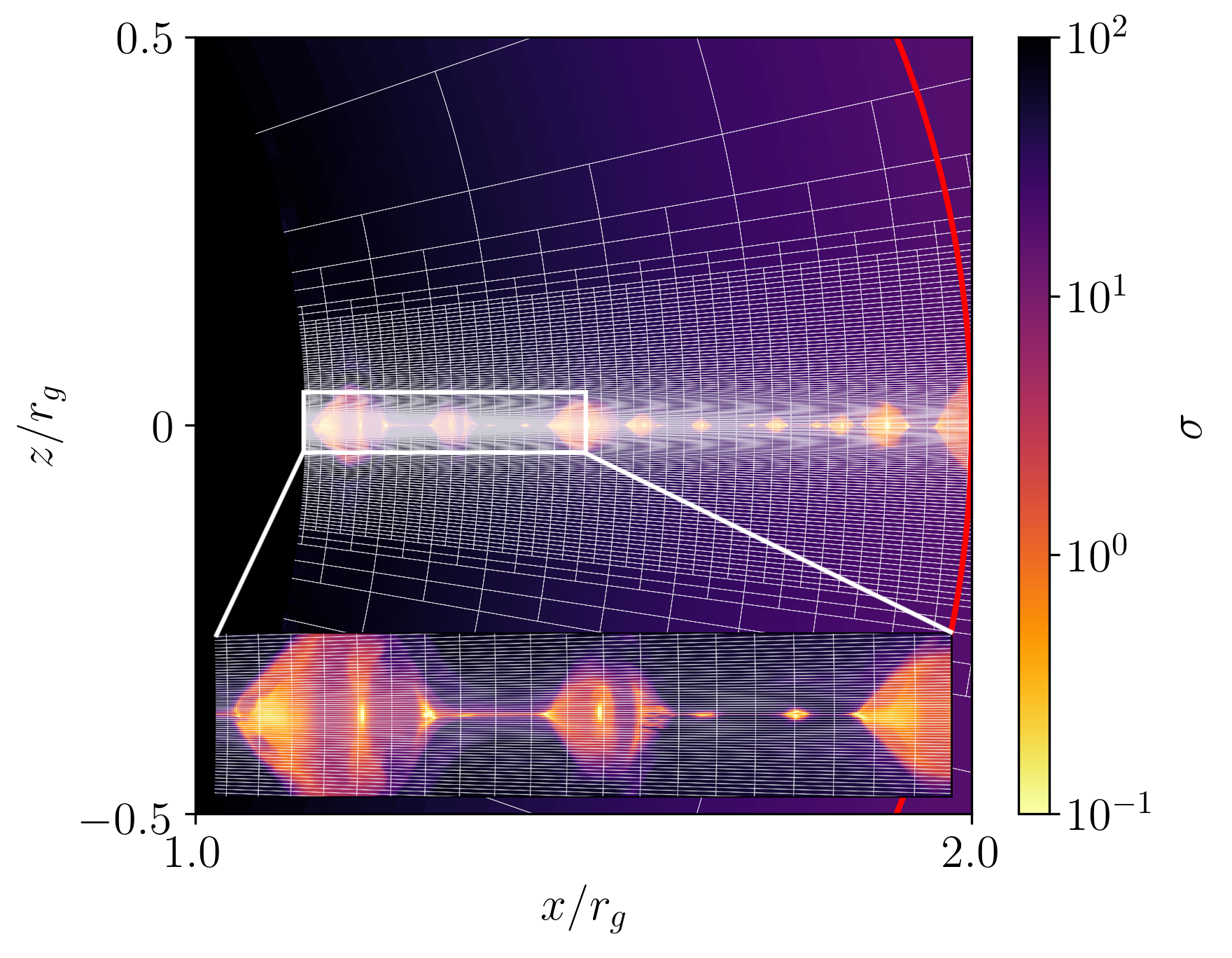} 
\caption{AMR blocks used in the simulation GRRMHD1, with inset zoomed on the current sheet near the event horizon. Color shows magnetization $\sigma=B^2/(4\pi \rho c^2)$, and the red curve shows the ergosphere boundary.}
\label{amr}
\end{figure}

We confirm that the current sheet is resolved in Figure \ref{amr}, showing the AMR blocks for simulation GRRMHD1 (each block contains $N_r \times N_\theta=16\times 8$ cells). The reconnection rate and the thinning of the sheet in GRRMHD is converged when there are at least 10 grid cells over the width of the current sheet once it finished the thinning process \cite{Ripperda_2020}. In the simulation GRRMHD1 the current sheet width is resolved by slightly more than one AMR block inside the ergosphere, indicating that GRRMHD1 is converged (Figure~\ref{convergence}, right panel dark red curve), while GRRMHD1 at $1/2$ resolution and $1/4$ resolution are not converged (Figure~\ref{convergence}, right panel red and yellow curves), yet still resolve the current sheet by more than one cell. 

The combination of an implicit-explicit (IMEX) time-stepping scheme to capture fast reconnection dynamics \cite{ripperda2019b}, together with adaptive mesh refinement (AMR) capabilities of {\tt{BHAC}} to accurately resolve the smallest scales in the system allows us to study resistive reconnection and plasmoid formation in the GRRMHD simulations \cite{Ripperda_2020}. For the GRPIC simulations a grid uniform in $\cos\theta$ helps to concentrate resolution on the equatorial current sheet and ensure the plasmoid formation is resolved. 

The flux decay time scale converges in the GRPIC simulations with increasing $r_H/r_L$. The small-$r_L$ model (GRPIC1) displays a full spectrum of plasmoid size, while small plasmoids are not present in the large-$r_L$ model (GRPIC3), thus indicating a transition from many plasmoids to few. We also observe disruption of the current sheet by rapidly growing kink modes in the large-$r_L$ model (GRPIC3).

\section{Collisionality of the Reconnection Layer}
\begin{figure}[h!]
\centering
\includegraphics[width=0.42\textwidth]{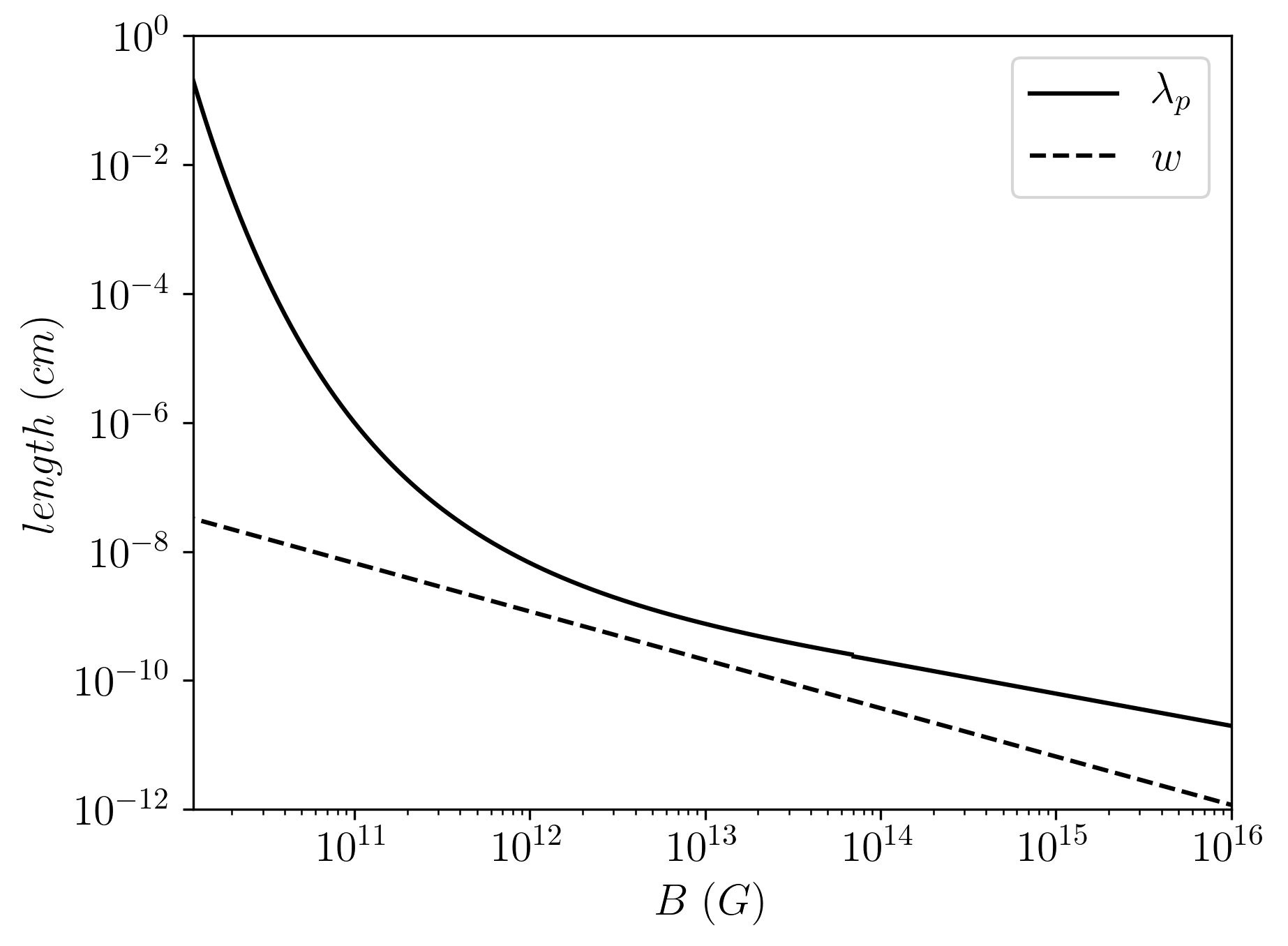} 
\caption{Elementary current-sheet width $w$ and plasma skin depth $\lambda_p$ vs magnetic field strength in the upstream, $B$.}
\label{collisional}
\end{figure}

The reconnection is collisionless when the plasma skin depth $\lambda_{p}$ is larger than the elementary current-sheet width in the resistive-MHD chain $w\sim 100 \eta/v_A\sim 100 \eta/c$ \cite{Bhattacharjee_2009,Uzdensky_2010}, where $\eta$ is the diffusivity due to coulomb collisions of pairs. Here we estimate analytically when this condition is satisfied. Following \cite{Uzdensky_2011}, we assume that the pressure of electron-positron pairs and radiation balances magnetic pressure in the reconnection layer,
\beq
P_\text{rad} + P_\text{pairs} = \frac{B^2}{8\pi},
\label{pressure}
\eeq
where $P_\text{rad}=(1/3)a T^4$ is the radiation pressure, $P_{\text{pairs}}$ is the pressure of the pair plasma, $B$ is the upstream magnetic field strength, and $a$ is the radiation constant. When $k_B T \ll m c^2$, radiation pressure dominates the LHS of Eq.~\ref{pressure}. At high temperatures $k_B T \gg m c^2$, $P_\text{pairs}\approx (7/4) P_\text{rad}$. We set $P_\text{pairs}=(7/4) P_\text{rad}$, which gives the temperature of the reconnnection layer
\beq
\frac{k_B T}{m c ^2} \approx 0.3~B_{12}^{1/2}.
\eeq
It is a reasonable estimate of $T$ in the non-relativistic and ultra-relativistic  regimes. Then the Spitzer diffusivity due to coulomb collisions of pairs is given by
\beq
\eta = \frac{c^2}{4\pi}\frac{4\sqrt{2\pi}e^2 m^{1/2}\text{ln}\Lambda }{3(k_B T)^{3/2}}\approx 0.4~B_{12}^{-3/4}~\text{cm}^2\text{s}^{-1},
\eeq
where we have set the coulomb logarithm $\text{ln}\Lambda=21$ \cite{Uzdensky_2011}. It determines the elementary current sheet width
\beq
w = 100 \eta / c \approx 10^{-9}~B_{12}^{-3/4}~\text{cm}.
\eeq
To estimate the density of pairs we assume annihilation balance in the reconnection layer: $e^+ + e^- \rightleftharpoons \gamma + \gamma$. When the plasma is non-relativistic ($k_B T / m c^2 \ll 1$) it gives $n=2(k_B T / 2\pi \hbar^2)^{3/2}\exp(-mc^2 / k_B T)$, and the plasma skin depth follows as $\lambda_p = (mc^2/4\pi n e^2)^{1/2}$. When the plasma is ultra-relativistic ($k_B T / m c^2 \gg 1$) the density is given by $n = 1.202 (3/2 \pi^2)(k_B T / \hbar c)^{3}$, and the skin depth follows as $\lambda_p = (\langle \gamma \rangle mc^2/4\pi n e^2)^{1/2}$ where $\langle \gamma \rangle = 3 k_B T / m c^2$  is the average particle lorentz factor. The two curves for $\lambda_p$ join smoothly in the intermediate regime. Figure~\ref{collisional} shows $w$ and $\lambda_p$ as a function of $B$. It suggests that the reconnection is collisionless when $B\ll 10^{12}$~G. However, a self consistent numerical calculation including detailed pair production and collisional physics is required to determine the reconnection rate if the magnetic field is very strong $B\gtrsim 10^{12}$~G, or if pair production is very efficient. 

\section{Magnetic Flux Decay Timescale}
The flux decay can be understood analytically by assuming a constant reconnection rate at the stagnation surface, and neglecting plasmoid formation. The integral form of Faraday's law gives
\beq
\frac{d\Phi}{dt} = -2\pi c E_\phi,
\label{Faraday}
\eeq
where we have evaluated the line integral on a circle of radius $r_\text{0}$ at $\theta=\pi/2$ (the stagnation surface on the equator). The flux is given by 
\beq
\Phi = \frac{1}{2}\int_0^{\pi} \int_0^{2\pi} |B^r|\sqrt{\gamma}d\theta d\phi,
\eeq
where $\sqrt{\gamma}$ is the spatial metric determinant. In this toy model the global flux decay is determined the local electric field $E_\phi$ at the equator on the stagnation surface. We make the ideal MHD approximation $E_\phi = \sqrt{\gamma}(v^\theta B^r - v^r B^\theta)/c = \sqrt{\gamma} v^\theta B^r /c$, since $v^r=0$ on the stagnation surface. Converting the velocity to a physical (tetrad) component $v^{\hat{\theta}} = \sqrt{g_{\theta \theta}} v^\theta $, gives 
\beq 
E_\phi = \sqrt{\gamma} \frac{v^{\hat{\theta}} B^r}{r_\text{0} c}.
\label{E_ph}
\eeq
Assuming uniform $B^r(\theta,r_0)$ would give $\Phi \approx \mathcal{S}(r_{0})B^r$, where $\mathcal{S}(r_{0})$ is the hemisphere area at radius $r_{0}$. However, the distribution of $B^r(\theta)$ is non-uniform and concentrated near the poles, similar to the asymptotic pulsar wind in Minkowski spacetime \cite{Michel_1973}. This is because our simulations begin with a dipole field which is twice is strong at the poles compared to the equator. The dipole field is then opened, resulting in a non-uniform split-monopole \cite{Tchekhovskoy_2016}. Therefore, we make the more precise statement $\Phi = k \mathcal{S}(r_{0}) B^r(r_0,\pi/2)$, where we measure $k\approx 1.7$. Combining this result with Equation~\ref{Faraday} and Equation~\ref{E_ph} gives
\begin{equation}
    \frac{d\Phi}{dt}=-\frac{2\pi \sqrt{\gamma} v^{\hat{\theta}}}{r_{0} k\mathcal{S}(r_{0})}\Phi.
\end{equation}
It implies exponential decay of $\Phi$ on the timescale 
\begin{equation}
    \tau = \frac{r_{0} k\mathcal{S}(r_{0})}{2\pi \sqrt{\gamma} v^{\hat{\theta}}},
    \label{tau}
\end{equation}
similar to the result of \cite{crinquand2020synthetic}. Equation~\ref{tau} gives $\tau \approx 3 r_g/ v^{\hat{\theta}} $ for $r_0\approx 2r_g$.

\section{Luminosity}
\begin{figure}[h!]
\centering
\includegraphics[width=0.48\textwidth]{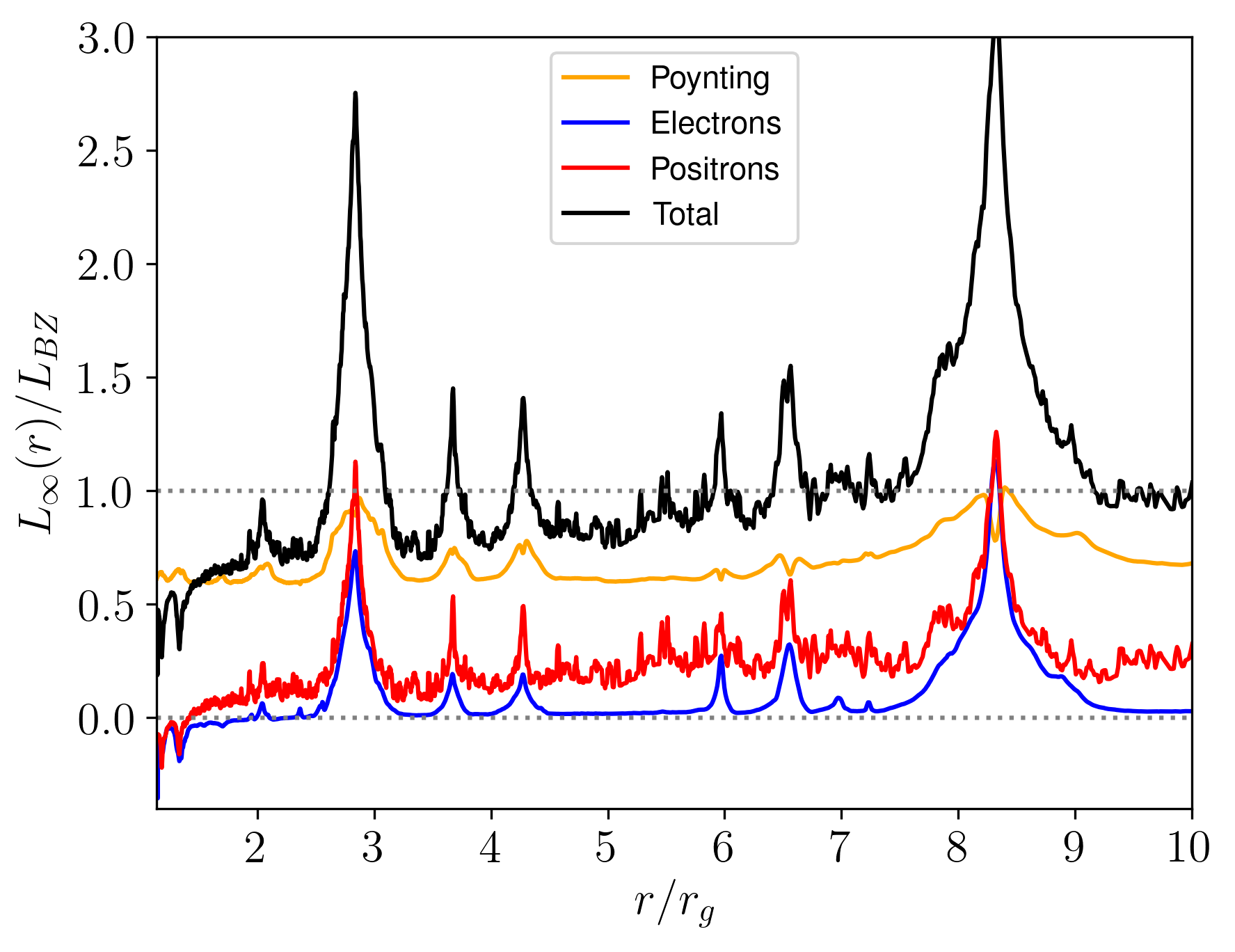} 
\includegraphics[width=0.48\textwidth]{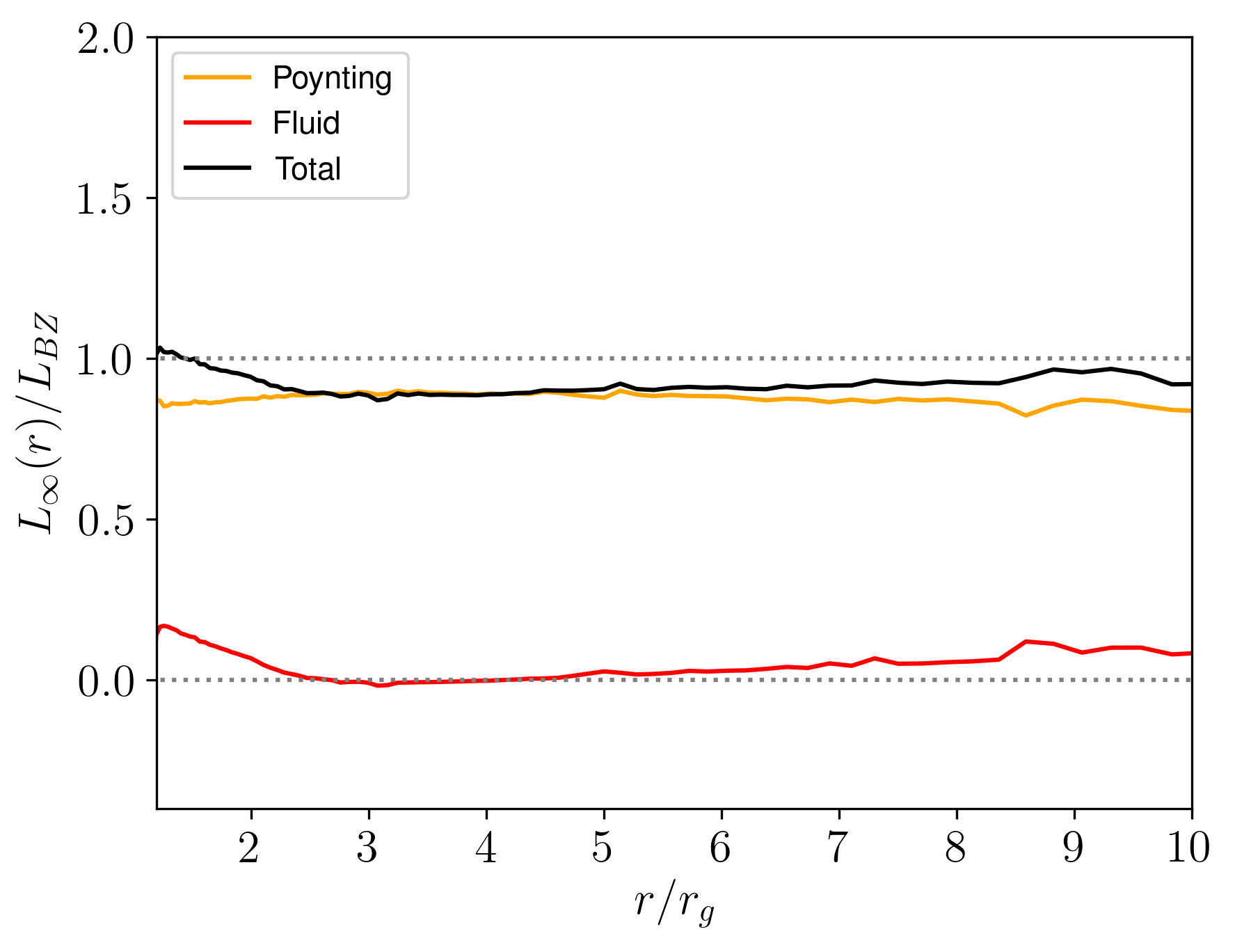} 
\caption{Flux of conserved energy through spherical shells of radius $r$, as seen by an observer at infinity in units of $L_\text{BZ}$. Left panel: GRPIC1 at $t=100$~$r_g/c$. Right panel: GRRMHD1 at $t=311$~$r_g/c$.}
\label{luminosity}
\end{figure}

Figure~\ref{luminosity} shows the flux of conserved energy passing through spherical shells of radius $r$ as seen by an observer at infinity. The flux is normalized to units of
\beq
L_\text{BZ}=\kappa\frac{\Omega_{H}^2\Phi^2}{4\pi c},
\eeq
where $\kappa=0.053$ for a split monopole \cite{Tchekhovskoy_2011}. The luminosity is dominated by the integrated Poynting flux, except in plasmoids in GRPIC simulations, where the particle luminosity can meet or exceed that of the fields. The particle luminosity is dominated by positrons in the GRPIC simulations because a net positive charge is required to support the rotating split-monopole with $\vec{\Omega}\cdot\vec{B}_p>0$ in both hemispheres. The particle luminosity in the ergosphere is mainly due to inward going positive energy particles. Large deviations in Poynting flux are seen at the location of plasmoids in GRPIC, while in GRRMHD the deviations are much smaller because of the smaller plasmoid sizes. 

\bibliography{Astrophysics}

\end{document}